\documentstyle[prl,epsf,amssymb,amsbsy,amstext,amsfonts,aps]{revtex}
\begin{document}
\twocolumn
\draft{}
\bibliographystyle{try}

\topmargin 0.0001cm

 
 \newcounter{univ_counter}
 \setcounter{univ_counter} {0}

\addtocounter{univ_counter} {1} 
\edef\INFNGE{$^{\arabic{univ_counter}}$ } 

\addtocounter{univ_counter} {1} 
\edef\SACLAY{$^{\arabic{univ_counter}}$ } 

\addtocounter{univ_counter} {1} 
\edef\MSU{$^{\arabic{univ_counter}}$ }

\addtocounter{univ_counter} {1} 
\edef\ASU{$^{\arabic{univ_counter}}$ } 

\addtocounter{univ_counter} {1} 
\edef\CMU{$^{\arabic{univ_counter}}$ } 

\addtocounter{univ_counter} {1} 
\edef\CUA{$^{\arabic{univ_counter}}$ } 

\addtocounter{univ_counter} {1} 
\edef\CNU{$^{\arabic{univ_counter}}$ } 

\addtocounter{univ_counter} {1} 
\edef\UCONN{$^{\arabic{univ_counter}}$ } 

\addtocounter{univ_counter} {1} 
\edef\DUKE{$^{\arabic{univ_counter}}$ } 

\addtocounter{univ_counter} {1} 
\edef\FIU{$^{\arabic{univ_counter}}$ } 

\addtocounter{univ_counter} {1} 
\edef\FSU{$^{\arabic{univ_counter}}$ } 

\addtocounter{univ_counter} {1} 
\edef\GWU{$^{\arabic{univ_counter}}$ } 

\addtocounter{univ_counter} {1} 
\edef\ORSAY{$^{\arabic{univ_counter}}$ } 

\addtocounter{univ_counter} {1} 
\edef\ITEP{$^{\arabic{univ_counter}}$ } 

\addtocounter{univ_counter} {1} 
\edef\INFNFR{$^{\arabic{univ_counter}}$ }

\addtocounter{univ_counter} {1} 
\edef\JMU{$^{\arabic{univ_counter}}$ } 

\addtocounter{univ_counter} {1} 
\edef\KYUNGPOOK{$^{\arabic{univ_counter}}$ } 

\addtocounter{univ_counter} {1} 
\edef\MIT{$^{\arabic{univ_counter}}$ } 

\addtocounter{univ_counter} {1} 
\edef\UE{$^{\arabic{univ_counter}}$ } 

\addtocounter{univ_counter} {1} 
\edef\UMASS{$^{\arabic{univ_counter}}$ }

\addtocounter{univ_counter} {1} 
\edef\UNH{$^{\arabic{univ_counter}}$ } 

\addtocounter{univ_counter} {1} 
\edef\NSU{$^{\arabic{univ_counter}}$ } 

\addtocounter{univ_counter} {1} 
\edef\OHIOU{$^{\arabic{univ_counter}}$ } 

\addtocounter{univ_counter} {1} 
\edef\ODU{$^{\arabic{univ_counter}}$ } 

\addtocounter{univ_counter} {1} 
\edef\PITT{$^{\arabic{univ_counter}}$ } 

\addtocounter{univ_counter} {1} 
\edef\RPI{$^{\arabic{univ_counter}}$ } 
 
\addtocounter{univ_counter} {1} 
\edef\RICE{$^{\arabic{univ_counter}}$ } 

\addtocounter{univ_counter} {1} 
\edef\URICH{$^{\arabic{univ_counter}}$ } 

\addtocounter{univ_counter} {1} 
\edef\SCAROLINA{$^{\arabic{univ_counter}}$ } 

\addtocounter{univ_counter} {1} 
\edef\UTEP{$^{\arabic{univ_counter}}$ } 

\addtocounter{univ_counter} {1} 
\edef\JLAB{$^{\arabic{univ_counter}}$ } 

\addtocounter{univ_counter} {1} 
\edef\VT{$^{\arabic{univ_counter}}$ } 

\addtocounter{univ_counter} {1} 
\edef\VIRGINIA{$^{\arabic{univ_counter}}$ } 

\addtocounter{univ_counter} {1} 
\edef\WM{$^{\arabic{univ_counter}}$ } 

\addtocounter{univ_counter} {1} 
\edef\YEREVAN{$^{\arabic{univ_counter}}$ }

\title{Photoproduction of the $\rho^0$ meson  on the proton at large momentum transfer}

 \author{ 
M.~Battaglieri,\INFNGE\
E.~Anciant,\SACLAY\
M.~Anghinolfi,\INFNGE\
R.~De Vita,\INFNGE\
E.~Golovach,\MSU\
J.M.~Laget,\SACLAY\
V.~Mokeev,\MSU\
M.~Ripani,\INFNGE\
G.~Adams,\RPI\
M.J.~Amaryan,\YEREVAN
% ...
\
\\(CLAS Collaboration)
}

\address{\INFNGE Istituto Nazionale di Fisica Nucleare, Sezione di Genova and Dipartimento di Fisica, Universit\`a di Genova, Italy 16146}
\address{\SACLAY CEA-Saclay, Service de Physique Nucleaire, Gif-sur-Yvette, France 91191}
\address{\MSU Moscow State University,Moscow, Russia 119899}
\address{\ASU Arizona State University, Tempe, Arizona 85287-1504}
\address{\CMU Carnegie Mellon University, Pittsburgh, Pennslyvania 15213}
\address{\CUA Catholic University of America, Washington, D.C. 20064}
\address{\CNU Christopher Newport University, Newport News, Virginia 23606}
\address{\UCONN University of Connecticut, Storrs, Connecticut 06269}
\address{\DUKE Duke University, Durham, North Carolina 27708-0305}
\address{\FIU Florida International University, Miami, Florida 33199}
\address{\FSU Florida State University, Tallahasee, Florida 32306}
\address{\GWU The George Washington University, Washington, DC 20052}
\address{\ORSAY Institut de Physique Nucleaire d'Orsay, IN2P3, BP 1, Orsay, France 91406}
\address{\ITEP Institute of Theoretical and Experimental Physics, Moscow, 117259, Russia}
\address{\INFNFR Istituto  Nazionale di Fisica Nucleare, Laboratori Nazionali di Frascati, Frascati, Italy 00044}
\address{\JMU James Madison University, Harrisonburg, Virginia 22807}
\address{\KYUNGPOOK Kyungpook National University, Taegu 702-701, South Korea}
\address{\MIT Massachusetts Institute of Technology, Cambridge, Massachusetts  02139-4307}
\address{\UE University of Edinburgh, Edinburgh, Scotland, UK EH9 3JZ}
\address{\UMASS University of Massachusetts, Amherst, Massachusetts  01003}
\address{\UNH University of New Hampshire, Durham, New Hampshirs 03824-3568}
\address{\NSU Norfolk State University, Norfolk, Virginia 23504}
\address{\OHIOU Ohio University, Athens, Ohio  45701}
\address{\ODU Old Dominion University, Norfolk, Virginia 23529}
\address{\PITT University of Pittsburgh, Pittsburgh, Pennslyvania 15260}
\address{\RPI Rensselaer Polytechnic Institute, Troy, New York 12180-3590}
\address{\RICE Rice University, Houston, Texas 77005-1892}
\address{\URICH University of Richmond, Richmond, Virginia 23173}
\address{\SCAROLINA University of South Carolina, Columbia, South Carolina 29208}
\address{\UTEP University of Texas at El Paso, El Paso, Texas 79968}
\address{\JLAB Thomas Jefferson National Accelerator Laboratory, Newport News, Virginia 23606}
\address{\VT Virginia Polytechnic Institute and State University, Blacksburg, Virginia   24061-0435}
\address{\VIRGINIA University of Virginia, Charlottesville, Virginia 22901}
\address{\WM College of Willliam and Mary, Williamsburg, Virginia 23187-8795}
\address{\YEREVAN Yerevan Physics Institute, Yerevan, Armenia 375036 }

\date{\today}

\maketitle

\newpage

\wideabs{
\begin{abstract}
The differential cross section, $d\sigma/dt$, for  $\rho^0$ 
meson photoproduction on the proton above the resonance 
region ($2.6<W<2.9$ GeV) was  measured
up to a momentum transfer $-t = 5$ GeV$^2$ 
using the CLAS detector
at the Thomas Jefferson National Accelerator Facility.
The $\rho^0$ channel was extracted from the measured two
charged-pion cross section by fitting the $\pi^+\pi^-$
and $p\pi^+$ invariant masses.
The low momentum transfer region shows 
the typical diffractive pattern expected from 
Reggeon-exchange. The flatter behavior at large $-t$ cannot be explained solely
in terms of QCD-inspired two-gluon exchange models.
The data indicate that  
other processes, like quark interchange, 
are important to fully
describe  $\rho$
photoproduction.
\end{abstract}

}

\narrowtext

We report the results of  $\rho^0$ meson photoproduction on protons 
for  $E_\gamma$  between 3.19 and 3.91 GeV. 
Data have been measured over the  $-t$ range from $0.1-5.0$ GeV$^2$
using the CEBAF Large Acceptance Spectrometer (CLAS)
at the Thomas  Jefferson National Accelerator Facility (TJNAF). 
The low momentum transfer region ($-t < 1$ GeV$^2$), already measured in previous 
experiments~\cite{Abbhhm,Ba72} at similar energies, shows a diffractive behavior
 interpreted in the framework of the Vector Meson Dominance (VMD) model~\cite{Ba78}
as the elastic scattering of  vector mesons off the proton target.
In a more recent approach,  this process is also described by the 
t-channel exchange of the Pomeron, scalar $\sigma$ meson~\cite{Fr96,Oh00},
and  $f_2(1270)$ Regge trajectories~\cite{La00}.

\begin{figure}[h]
\epsfxsize9.5cm
\epsfysize6.5cm
\centerline{\epsffile{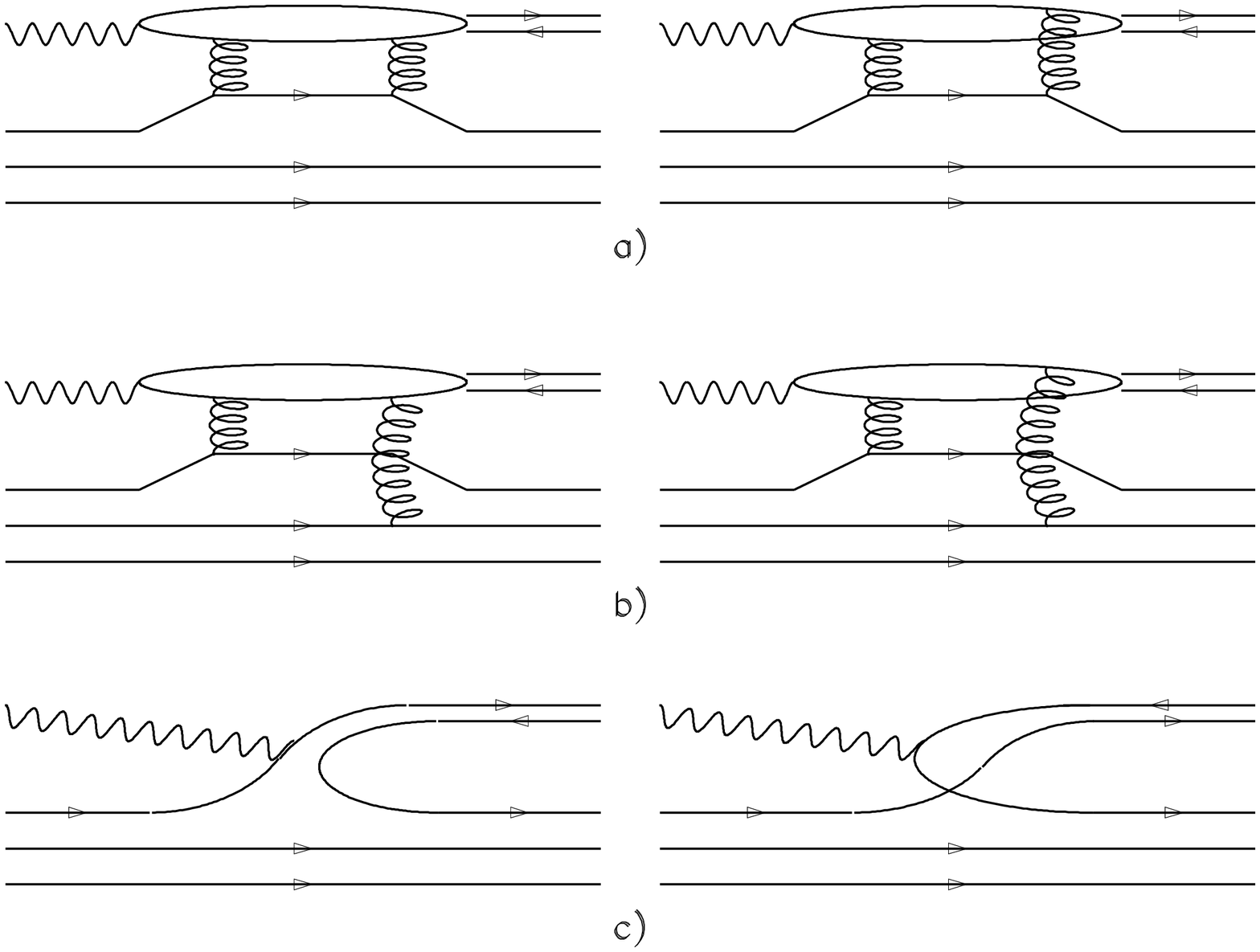}}
\caption[]{
The Feynman diagrams corresponding to 
a) two-gluon exchange from a single quark,
b) two-gluon exchange taking into account quark correlations in the nucleon,
and c) quark exchange.
}
\label{diagrams}
\end{figure}

At high $-t$, where the cross section  is sensitive to the microscopic details 
of the interaction, the underlying physics can be described  using 
parton degrees of freedom.
In a QCD-inspired framework,  the small impact
parameter ($\approx 1/{\sqrt{-t}}$)  prevents the constituent gluons (quarks) of the exchange
from interacting and forming a Pomeron (Reggeon).  Within certain models~\cite{DL87,La95}  this
means that the constituents can be resolved into two-gluon (two-quark)
structures (Fig.~\ref{diagrams}-a). 
Moreover, small transverse
sizes select configurations where each gluon couples to
different quarks both in the vector meson~\cite{La95} and the nucleon~\cite{La98}, giving access to 
the correlation function in the proton (Fig.~\ref{diagrams}-b)~\cite{La00}. 
The recent CLAS measurement of the $\phi$ photoproduction cross section~\cite{An00},
where the quark exchange is strongly suppressed by the OZI rule, was able
to isolate the contribution due to two-gluon exchange.
In the $\rho$ case, its light quark composition also allows  valence
quarks to be exchanged between the baryon  and the meson states
(Fig.~\ref{diagrams}-c)~\cite{La00}.

A bremsstrahlung photon beam was produced by an $E_0=4.1$ GeV continuous electron beam  hitting 
a gold
radiator of   $10^{-4}$ radiation lengths. The Hall B tagging system~\cite{SO99},
 with a photon energy resolution of 0.1\% E$_0$, 
was used to tag photons in the energy range from $3.19-3.91$ GeV.
The target cell, a mylar cylinder  6 cm in diameter and 18 cm long, was filled
with liquid hydrogen at 20.4 K.
During  data taking  at high intensity ($\sim 4\cdot 10^6 \gamma$/s),
the photon flux was continuously measured by an $e^+ e^-$  pair spectrometer located beyond the target. 
The efficiency of this device was determined during dedicated low intensity ($\sim 10^5 \gamma$/s )
runs  by comparison with  a 100\% efficient lead-glass total absorption counter.
The systematic uncertainty on the photon flux has been estimated to be 3\%~\cite{An00}.

The hadrons were detected in CLAS~\cite{Br00}, a spectrometer with nearly  
4$\pi$ coverage  that is based on a toroidal magnetic field ($\sim 1 T$) generated
by six superconducting coils. 
The field was set to bend the positive particles
away from the beam into the acceptance of the 
detector.
Three drift chamber regions allowed tracking of charged particles~\cite{DC} and
time-of-flight scintillators (TOF) were used for hadron
identification~\cite{Sm99}. The momentum resolution  was of the order of a few percent,
while the detector geometric acceptance  
was about 70\% 
for protons and positive pions.
Low energy negative particles, however, were mainly lost at 
forward angles because they were bent out of the acceptance.
Coincidences between the photon tagger and the CLAS detector (TOFs)
triggered the recording of hadronic interactions. From a total of  70M triggers,
0.8M events were identified as $p\pi^+\pi^-$ candidates.

The five-fold differential cross section 
\begin{equation} 
\frac{d\sigma}{d\tau} \equiv \frac{d \sigma(\gamma p \rightarrow  p \pi^+  \pi^-) }{dM_{\pi^+\pi^-}dM_{p\pi^+} d\phi^{cm}_{\rho}  d\phi^{decay}_{\pi^+} dt}
\end{equation}   
was measured by analyzing all possible event topologies of the reaction 
$\gamma p \rightarrow  p \pi^+  \pi^-$ with 
at least two detected hadrons in the final   state ($p \pi^+$, $p \pi^-$, $\pi^+\pi^-$,
$p \pi^+  \pi^-$), using the missing mass and missing energy techniques to eliminate
the underlying multi-pion background.

The CLAS acceptance and reconstruction efficiency  were evaluated with  Monte Carlo simulations. 
An event generator~\cite{Co94}  was used that contained the three main contributions to the $p \pi^+  \pi^-$
final state ($\gamma p \rightarrow  p \rho^0$, $\gamma p \rightarrow   \Delta^{++} \pi^-$,
 and $\gamma p \rightarrow  p \pi^+  \pi^-$ in $s$-wave), along with background reactions with 
three or more pions.
The generated events were processed by a GEANT-based code simulating the CLAS detector that 
reconstructed the simulated data using the same analysis procedure applied to the raw data.
To minimize the model dependence, the acceptance was derived as a
function of six independent kinematic variables describing the full three-body reaction, namely,
 $E_\gamma$,  the  two invariant masses $M_{p\pi^+}$ and 
$M_{\pi^+\pi^-}$, the  $\rho^0$ center-of-mass azimuthal angle $\phi^{cm}_{\rho}$, the $\pi^+$ 
azimuthal angle in the $\rho^0$ helicity frame $\phi^{decay}_{\pi^+}$,  and the momentum transfer $t$.
For each detected topology, the data were binned in the six dimensional space and corrected,
bin by bin, by the corresponding acceptance. 
Depending on the kinematics, 
the average acceptance of CLAS for these final states ranges from $5\%$ to $15\%$.

Each topology predominantly covers complementary kinematic regions.  
Combining all event topologies together, we were able to measure almost  the entire allowed phase
space. 
For $-t<0.1$ GeV$^2$ 
the CLAS detector had no acceptance  for
the reaction. 
When multiple topologies were present
in the same kinematic region, we considered 
each topology  as an independent measurement 
of the two pion cross section. 
Their comparison gave  
an evaluation of the systematic uncertainty on the cross section
ranging from 5\% to 10\%, depending on the kinematics.

To determine the relative weight of different channels that contribute 
to the $p \pi^+ \pi^-$ final state,
the two differential cross sections 
$d^2\sigma/dtdM_{\pi^+\pi^-}$ and $d^2\sigma/dtdM_{p\pi^+}$ 
were simultaneously fit to a phenomenological model~\cite{Mo99,Mo00,Ri00} that 
describes  two-pion production as a superposition of three-body phase space 
and quasi-two-body channels
($\gamma p \rightarrow p \rho^0$, 
$\gamma p \rightarrow p f_2(1270)$, 
$\gamma p \rightarrow  \Delta^{++} \pi^-$, 
$\gamma p \rightarrow  \Delta^{0} \pi^+$)
with subsequent decays. 
This model  describes the cross section as the sum of six
amplitudes:

\begin{equation} 
\frac{d\sigma}{d\tau} = \left\vert{{\sum_{i=1}^6{a_iT_i}}}\right\vert^2.
\end{equation} 

The complex reaction amplitudes $T_i$ correspond to: 
1)  diffractive processes in the $\gamma p \rightarrow p \rho^0$ reaction 
(Pomeron and  $f_2(1270)$ Regge trajectory exchanges); 
2) u-channel exchange in $\gamma p \rightarrow p \rho^0$ (nucleon trajectory);
3) Born terms in the $\gamma p \rightarrow  \Delta^{++} \pi^-$ and 
$\gamma p \rightarrow  \Delta^{0} \pi^+$ reactions (contact and $\pi$-in-flight) treated
in a Regge approach~\cite{Gu97};
4) $\rho^0$ Regge trajectory exchange in the reaction $\gamma p \rightarrow p f_2(1270)$;
5)  s-channel resonance excitation
for $p \rho^0$, $\Delta^{++} \pi^-$, and $\Delta^{0} \pi^+$ (20 well-established
resonances were included in the model even  though  the main contributions come from 
the $F_{35}(1905)$, $F_{37}(1950)$, and the $G_{17}(2190)$);
6) a phase space parameterized as a real constant number in each $(W,t)$ bin.
The $\rho^0$ and $\Delta^{++}$  decay amplitudes  were 
evaluated using an effective Lagrangian model with the form factors as given in Ref.~\cite{Long}.

\begin{figure}[h]
\epsfxsize9.cm
\epsfysize10.cm
\centerline{\epsffile{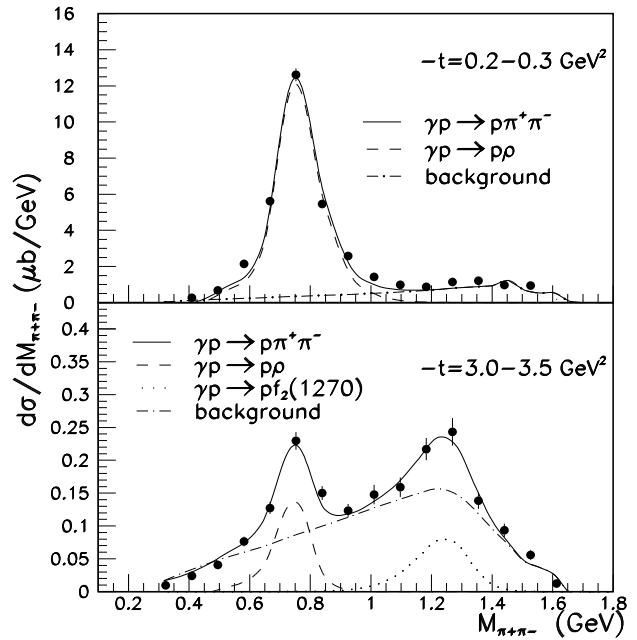}}
\caption[]{
The measured invariant mass distribution, $M_{\pi^+\pi^-}$, and the fitted quasi-two-body 
contributions in two $t$ bins: $0.2-0.3$ GeV$^2$ and $3.0-3.5$ GeV$^2$. 
The solid line corresponds to the coherent sum of different channels. 
}
\label{inv_masses}
\end{figure}

Figure~\ref{inv_masses} shows typical invariant mass distributions in
a low and a high $t$ bin, and the fitted 
decomposition into the two-body channels. 
The real fit parameters  $a_i$, determined in each $t$ bin 
using MINUIT~\cite{minuit}, show a smooth and small dependence on $t$,
indicating that  this model can  provide a reasonable description of the data. 

To evaluate the model dependence, 
the $\rho$ channel separation was also performed by means of 
a procedure similar to that used in the ZEUS data analysis~\cite{ZEUS},
fitting the $M_{\pi^+\pi^-}$ distribution with a third-order polynomial background 
plus Breit-Wigner-shaped lines and an interference term.
The extracted $\rho^0$ cross section agrees within 10\% to 30\%
with the results obtained using the phenomenological model.
A full description of the data analysis can be found in Ref.~\cite{Ba01}.

\begin{figure}[h]
\epsfxsize7.5cm
\epsfysize8.5cm
\centerline{\epsffile{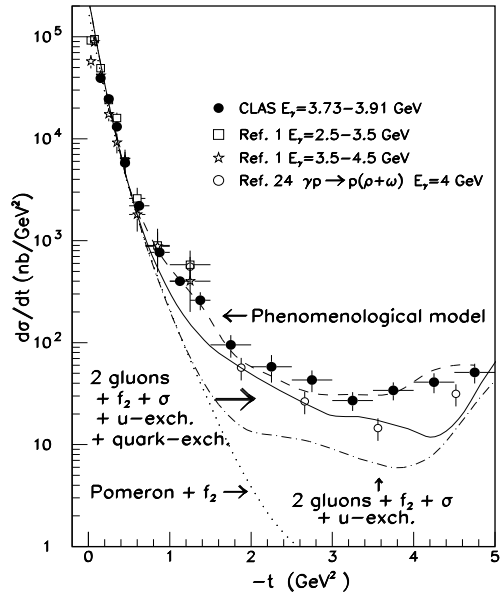}}
\caption[]{ The differential $\rho^0$ photoproduction cross section (see text for detailed explanation of the curves). 
The error bars include statistical and systematic uncertainties summed in quadrature.}
\label{ds_rho}
\end{figure}

The extracted $\rho^0$ photoproduction cross section as a function of $t$
is presented in Fig.~\ref{ds_rho} 
for  $E_\gamma=3.8$ GeV.
In the low momentum transfer region 
good agreement with the previous measurement of Ref.~\cite{Abbhhm} is evident.
Assuming an exponential $A e^{Bt}$  
behavior in the range  $0.1<-t<0.5$ GeV$^2$, the  coefficient resulting from this experiment,
$B =-6.4\pm0.3$ GeV$^{-2}$,  is consistent with the value  $B =-6.9\pm0.4$ GeV$^{-2}$ 
quoted in Ref.~\cite{Abbhhm}. 
The existing  data at large momentum transfer were taken at SLAC~\cite{An76}  
with a bremsstrahlung photon beam  and  a single arm spectrometer; the signal was 
unfolded from the background using a Breit-Wigner fit but 
the incomplete detection  of the final state did not allow separation of the $\omega$ 
channel from the $\rho^0$ channel, the latter comprising 50-75\% of the data set.

Predictions from presently available models are also shown in Fig.~\ref{ds_rho}.
The dashed line corresponds to the phenomenological model
used for the channel separation assuming  that
the  parameters $a_i$ of equation $(2)$ are
independent of $t$.
In this model, the  Pomeron and $f_2(1270)$ Regge 
trajectory exchanges (dotted line) describe the low momentum transfer region, 
while  the large $-t$ flat behavior is reproduced by the tails of 
resonances having a sizeable branch  in the $\rho$ channel.  
The overall resonance coupling was found to be compatible
with what is expected by  extrapolating from the low $W$ region by
means of a Breit-Wigner shape with an
energy-dependent width. 
 
In the QCD-inspired model of Refs.~\cite{La00,La98} (dot-dashed line on Fig.~\ref{ds_rho}),
the Pomeron exchange has been  replaced by the exchange of two 
non-perturbatively dressed gluons.
At low $-t$ the good agreement with the data is obtained adding the $f_2(1270)$ and the 
$\sigma$ Regge trajectories, while the rise at large $-t$ is given by the 
Reggeized u-channel exchange ($N$ and $\Delta$ trajectories)~\cite{La00}.
In this model the gluons can couple to any quark in the $\rho$ 
meson and in the  baryon (see Fig.~\ref{diagrams}-a,b) and 
quark  correlations  in the proton are taken into account
assuming the simplest form of its wave function with three valence quarks 
equally sharing the proton  longitudinal  momentum~\cite{Cu94}.

\begin{figure}[h]
\epsfxsize7.5cm
\epsfysize8.5cm
\centerline{\epsffile{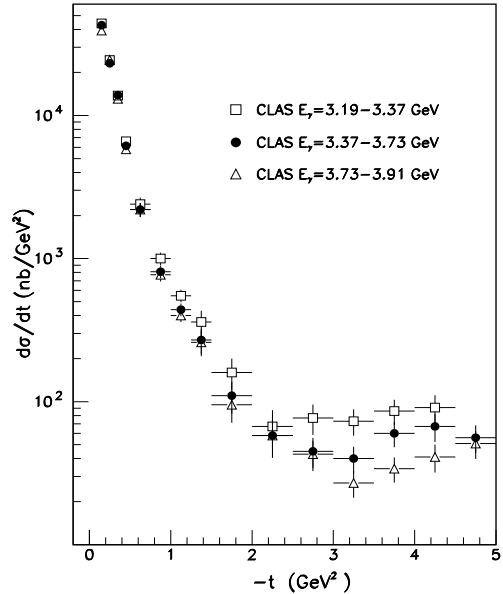}}
\caption[]{ The energy dependence of the $\rho^0$ photoproduction $d\sigma/dt$ as measured in CLAS.}
\label{rho_w_dep}
\end{figure}

The QCD-inspired model describes  $\phi$ photoproduction over the entire $t$ range in our energy
region~\cite{An00}, as well as the higher energy $\rho^0$
SLAC results ($E_\gamma\sim$ 6 GeV)~\cite{La00} where two-gluon exchange is also expected 
to dominate.
Instead, as shown in Fig.~\ref{ds_rho}, the model  underestimates the CLAS data,
leaving  room for quark-exchange processes (Fig.~\ref{diagrams}-c).
As explained in Refs.~\cite{Gu97,Br73,Co84} these hard-scattering mechanisms can be
incorporated in an effective  way by using the so called ``saturated'' trajectory.
Regge trajectories are usually linear in $t$ but are expected  to ``saturate'',
i.e. be $t$-independent, at large momentum transfer~\cite{Se94}. 
The trajectory has been chosen as \mbox{$\alpha(t)=-1$} when $-t>3$ GeV$^2$ ~\cite{La00} 
to be compatible with quark counting rules~\cite{Br73}.

The solid line in Fig.~\ref{ds_rho} shows the full calculation,
including such  saturating trajectory. Quark exchange  
increases the cross section at large $-t$ by a factor two.

Figure~\ref{rho_w_dep} shows the measured cross sections in different photon energy bins.
From the three data sets, the $\rho^0$ cross section at
$\theta_{cm}=90^\circ$ is extracted as a function of energy.
The power law  $s^{-C}$ fit to $d\sigma/dt$ at $90^\circ$ in the center of mass is
performed using  both SLAC~\cite{Ba72,An76} and CLAS data.
The fit yields $C=7.9\pm0.3$ ($\chi^2=0.6$)  showing a good agreement with    $s^{-8}$ behavior.
The quark exchange diagrams 
of Fig.~\ref{diagrams}-c-left (point-like interaction) and~\ref{diagrams}-c-right
(hadronic component of the photon)  have
a $s^{-7}$ and $s^{-8}$  power-law behavior respectively, both by dimensional counting~\cite{Br73} and 
by recent models~\cite{Ha00}.
Note that also the saturated $\sigma$ Regge trajectory behaves like $s^{-8}$.
Like the differential cross section at fixed energy, 
the $s$ dependence suggests the presence of  quark interchange
hard mechanisms in addition to the  two-gluon exchange process.

In conclusion, the full differential two charged-pion
cross section was measured for the first time over a large
angular range. 
The differential cross section for $\gamma p\to p \rho^0$  was derived by
fitting the  $M_{\pi^+\pi^-}$ and $M_{p\pi^+}$ mass distributions 
with a realistic phenomenological model.
The comparison 
with available  models 
provides  indications of the presence of hard processes.
Adopting  a QCD language in this energy region, the two-gluon exchange mechanisms
(that fully  describe the $\phi$ photoproduction data) are not sufficient 
to reproduce the cross section at large momentum transfer
and its energy dependence.
Good agreement is achieved when 
quark interchange processes (suppressed in $\phi$ production) are  included
in an effective way 
in the calculation for the $\rho^0$ production.

We would like to acknowledge the outstanding efforts of the staff of the Accelerator
and the Physics Divisions at JLab that made this experiment possible. 
This work was supported in part by  the  Italian Istituto Nazionale di Fisica Nucleare, 
the French Centre National de la Recherche Scientifique and Commissariat \`{a} l'Energie Atomique, 
the U.S. Department of Energy and National Science Foundation, 
and the Korea Science and Engineering Foundation.
The Southeastern Universities Research Association (SURA) operates the
Thomas Jefferson National Accelerator Facility for the United States
Department of Energy under contract DE-AC05-84ER40150.


\begin{thebibliography}{99}
    \bibitem{Abbhhm} ABBHHM Collaboration, Phys. Rev. {\bf 175},  1669 (1968).
    \bibitem{Ba72}  J. Ballam {\it et al.}, Phys. Rev. {\bf D5},  545 (1972).
    \bibitem{Ba78} T.H. Bauer {\it et al.}, Rev. Mod. Phys. {\bf 50}, 261 (1978).
    \bibitem {Fr96} B. Friman and M. Soyeur, Nucl. Phys.  {\bf A600}, 477  (1996).
    \bibitem {Oh00} Y. Oh, A.I. Titov, T.-S.H. Lee, Proceedings of the NSTAR2000 Conference 
    edited by V. Burkert, L. Elouadrhiri, J. Kelly, R. Minehart, {\it World Scientific}, p. 255. 
    \bibitem {La00} J.M. Laget,   Phys. Lett.  {\bf 489B}, 313 (2000).

     \bibitem{DL87}  A. Donnachie and P.V. Landshoff, Phys. Lett. {\bf B185}, 403 (1987).
    \bibitem {La95} J.M. Laget and R. Mendez-Galain, Nucl. Phys.  {\bf A581}, 397  (1995).
     \bibitem{La98}  J.M. Laget,  {\it Physics and Instrumentation with 6--12 GeV Beams}, edited by 
                     S. Dytman, H. Fenker, and P. Ross (Jefferson Lab User Production, 1998), p. 57.
    \bibitem{An00}  E. Anciant  {\it et al.}, Phys. Rev. Lett. {\bf 85},  4682 (2000).
     \bibitem{SO99}  D.I. Sober {\it et al.}, Nucl. Instr. and Meth. {\bf A440}, 263  (2000).
    \bibitem {Br00} W. Brooks, Nucl. Phys.  {\bf A663-664}, 1077  (2000).
    \bibitem{DC}    D. Carman    {\it et al.}, Nucl. Instr. and Meth. {\bf A449}, 81 (2000).
    \bibitem{Sm99}  E.S. Smith {\it et al.}, Nucl. Instr. and Meth. {\bf A432}, 265  (1999).
    \bibitem{Co94}  P. Corvisiero {\it et al.}, Nucl. Instr. and Meth. {\bf A346}, 433  (1994).
    \bibitem{Mo99}  V. Mokeev  {\it et al.}, Few Body Syst. Suppl. {\bf 11}, 292 (1999).
    \bibitem{Mo00}  M. Ripani  {\it et al.}, Phys. of Atom. Nucl. {\bf 63}, 1943 (2000).
    \bibitem{Ri00}  M. Ripani {\it et al.},  Nucl. Phys.  {\bf A672}, 220  (2000).
    \bibitem {Gu97} M. Guidal, J.M. Laget, and M. Vanderhaeghen, Nucl. Phys.  {\bf A627}, 645  (1997).
    \bibitem{Long}  R.B.Longacre and J.Dolbeu,  Nucl. Phys.  {\bf B122}, 493 (1977).
    \bibitem{minuit}  F.James CERN Program Library D506. 
    \bibitem{ZEUS}  The ZEUS Collaboration  Eur. Phys. J.  {\bf C14}, 213 (2000).
    \bibitem{Ba01}  M. Battaglieri, CLAS-Analysis 2001-002 May 2001 
     \texttt{http://www.jlab.org/Hall-B/pubs/\\ 
     \mbox{$\;\;\;\;\;\;\;\;\;\;\;$ analysis/battaglieri\_rho.ps}}
     \bibitem{An76}   R.L. Anderson {\it et al.}, Phys. Rev. {\bf D14},  679 (1976). 
    \bibitem {Cu94} J.R. Cudell and B.U. Nguyen, Nucl. Phys.  {\bf B420}, 669  (1994).    
    \bibitem{Br73}  S.J. Brodsky and G.R. Farrar, Phys. Rev. Lett. {\bf 31},  1153 (1973).
    \bibitem{Co84}  P.D.B Collins and P.J. Kearney, Z. Phys. {\bf C22},  277 (1984).
    \bibitem{Se94}  M.N. Sergeenko, Z. Phys. {\bf C64},  315 (1994).
    \bibitem{Ha00} H.W Huang and P. Kroll, Eur. Phys. J. {\bf C17}, 423   (2000).

\end{thebibliography}
\end{document}